# Free-surface flow simulations for discharge-based operation of hydraulic structure gates

C.D. Erdbrink[1,2,3], V.V. Krzhizhanovskaya[1,2], P.M.A. Sloot[1,2]

[1] *University of Amsterdam, Amsterdam, The Netherlands*
[2] *University of Information Technologies, Mechanics and Optics, Saint Petersburg, Russia*
[3] *Deltares, Delft, The Netherlands*

**Abstract**

We combine non-hydrostatic flow simulations of the free surface with a discharge model based on elementary gate flow equations for decision support in operation of hydraulic structure gates. A water level-based gate control used in most of today's general practice does not take into account the fact that gate operation scenarios producing similar total discharged volumes and similar water levels may have different local flow characteristics. Accurate and timely prediction of local flow conditions around hydraulic gates is important for several aspects of structure management: ecology, scour, flow-induced gate vibrations and waterway navigation. The modelling approach is described and tested for a multi-gate sluice structure regulating discharge from a river to the sea. The number of opened gates is varied and the discharge is stabilized with automated control by varying gate openings. The free-surface model was validated for discharge showing a correlation coefficient of 0.994 compared to experimental data. Additionally, we show the analysis of CFD results for evaluating bed stability and gate vibrations.

**Notation**

| | |
|---|---|
| A | amplitude of gate vibration (m) |
| $A_{lake}$ | surface area of lake (m$^2$) |
| a | gate opening (m) |
| $C_c$ | contraction coefficient for flow past sharp-edged underflow gate (-) |
| $C_{c,in}$ | contraction coefficient for flow entering upstream section between piers (-) |
| $C_d$ | discharge coefficient for submerged flow (-) |
| $C_E$ | discharge coefficient as used by Nago(1983) for experimental data |
| e | error value of PID discharge controller (m$^3$/s) |
| $f_{gate}$ | frequency of gate vibration (Hz) |
| Fr | Froude number (-) |
| g | gravitational constant (m$^2$/s) |
| h | water depth (m) |
| $h_0$ | upstream water depth before reaching pier (m) |
| $h_1$ | upstream water depth between piers, upstream of gate (m) |
| $h_2$ | water depth in control section (m) |
| $h_3$ | water depth downstream of gate, between piers, behind recirculation zone (m) |
| $h_4$ | downstream water depth beyond pier (m) |
| $h_{target}$ | target lake level to be reached at the end of discharge period (m) |
| k | turbulent kinetic energy (m$^2$/s$^2$) |
| $K_P, K_I, K_D$ | gain parameters of PID discharge controller (-) |
| m | number of gates opened fully or partially (-) |
| n | total number of gates of the structure (-) |
| $\vec{n}$ | normal vector (-) |
| p | pressure (Pa) |
| Q | discharge (m$^3$/s) |
| $Q_{gate,i}$ | discharge through gate *i* (m$^3$/s) |
| $Q_{MF}$ | modular flow discharge based on gate underflow contraction criterion (m$^3$/s) |
| q | discharge per unit width (m$^2$/s) |
| t | time variable (s) |
| U | magnitude of flow velocity vector; in 2DV model defined as $U = \sqrt{u^2 + w^2}$ (m/s) |
| $U_{vc}$ | magnitude of flow velocity vector in vena contracta (m/s) |
| (u, w) | flow velocity vector in 2DV model; *u* is horizontal velocity, *w* is vertical velocity (m/s) |
| Vr | reduced velocity parameter of flow-induced vibrations (-) |
| $V_{tot}$ | total volume passing the structure in a given amount of time (m$^3$) |
| $V_{tot,req}$ | required total volume to pass the structure in a given amount of time in order to reach $h_{target}$ (m$^3$) |
| w | width between piers (m) |
| α | calibration parameter for turbulent flow in bed stability parameter (-) |
| ε | turbulent dissipation (m$^2$/s$^3$) |
| $\xi_{in}$ | entrance loss coefficient (-) |
| $\xi_{out}$ | exit loss coefficient (-) |
| Ψ | stability parameter for beginning of motion of granular bed material (-) |
| $\binom{q}{r}$ | number of combinations of *r* objects out of *q* objects (0 ≤ *r* ≤ *q*), defined as $\frac{q!}{r!(q-r)!}$ |
| $\lfloor r \rfloor$ | floor function, defined as $\forall r \in \mathbb{R}, \lfloor r \rfloor = \max(n \in \mathbb{Z}: n \leq r)$ |
| $\bar{r}$ | time-average of quantity *r* |

**Abbreviations**

| | |
|---|---|
| 2DV | two-dimensional model in the vertical |
| ALE | arbitrary Lagrangian-Eulerian |
| CFD | computational fluid dynamics |
| FEM | Finite Element Method |
| FIV | flow-induced vibrations |
| PID | proportional integral derivative controller |
| TKE | turbulent kinetic energy |
| RANS | Reynolds-Averaged Navier-Stokes |

## INTRODUCTION

This paper gives an outline of how near-field free-surface flow simulations can be used in the operation of gates of large hydraulic structures.

Barrier operation is commonly based on water level predictions from system-scale far-field flow models. The procedures are aimed at fulfilling the main function of the structure: for a weir in a river this is to maintain the upstream water level; for a discharge sluice this is to transfer river water out to the sea while keeping a safe inland level. Present-day hydraulic structures have various secondary functions, such as providing favourable ecological conditions, for which usually no numerical aids are available in daily operation. A better prediction of the flow near structures would be beneficial to durable performance of all barrier tasks.

Proper design studies pay attention to all functions of a structure and assess the impact of all relevant flow features. However, operational constraints change in time for natural reasons (e.g. sea-level rise) or political reasons (e.g. "Kierbesluit Haringvlietsluizen", see Rijkswaterstaat 2004). In addition, sometimes the design criteria that were originally applied cannot be retrieved, yielding uncertainty about safety levels and allowable limits of gate settings in the present.

There are several aspects in contemporary barrier management for which an informed view on discharge and flow properties around gates is essential:
- stability of bed material and scour prediction: local erosion behind the bed protection (Hoffmans & Pilarczyk 1995) as well as larger scale morphological changes of surrounding bathymetry (Nam et al. 2011);
- ecological issues: fish migration, salt water intrusion and mobile fauna (Martin et al. 2005);
- dynamic forces associated with flow-induced gate vibrations (Naudascher & Rockwell 1994);
- impact of flow around structure on nearby shipping traffic;
- abnormal conditions: downtime of gates during scheduled maintenance or unexpected gate failure;

The barriers and sluices built in the South-West of The Netherlands in the period 1960-2000 are good examples of structures where different functions are combined. Present management of the barriers at Haringvliet and Oosterschelde calls for smart use to allow for regulation of fresh and salt water flows and fish migration. At the same time the aging process of these structures demands an increasing awareness on structural safety issues. The new storm-surge barrier of Saint Petersburg, Russia, is another example. This large dam houses two sector-gates and three sections of radial gates that protect the low-lying city centre and regulate the discharge from the river Neva. Operation of this complex structure must rely on state-of-the-art flow models.

Above considerations motivate quantification of flow around a hydraulic structure. The aim of this paper is to lay the foundation of numerical models to estimate gate discharges and evaluate the impact of flow near hydraulic structures in a way that is fit for operational applications. The influence of waves is not investigated; the focus is on flow (currents).

Traditionally, flow around hydraulic structures is studied experimentally in the design stage or as fundamental research topic (Roth & Hager 1999; Kolkman 1994). Numerous numerical studies have looked into sluice gate flow (Akoz et al. 2009; Kim 2007; Khan et al. 2005), but no single accepted, validated modelling tool exists for assessing turbulent gate flow with suitable practical value. Estimating discharge over weirs or under gates is not trivial. New discharge equations are still being introduced, both from informatics viewpoints (Khorchani & Blanpain 2005) and from traditional viewpoint of measurements (Habibzadeh et al. 2011).

System-scale models of inland water systems simulate the flow in river branches by solving the one-dimensional or quasi-two dimensional Shallow Water equations (Deltares 2012a, b). The fact that these hydrostatic models do not simulate the flow around hydraulic structures explicitly is not a severe limitation for most applications. The system effect of the operation of various gates on the water levels in adjacent water bodies (river branches) can be studied, for instance (Becker &

Schwanenberg 2012). For stability of granular bed material and salt water transport, however, the flow acceleration in the vertical dimension needs to be simulated. Moreover, the downside of primarily water level-centred validation and calibration in combination with parameterized structure representations (constant discharge coefficients, for example) is that the prediction quality of discharges in system-scale models is often unclear. Warmink et al. (2007, 2008) investigated the uncertainty in calibration of water levels in river models resulting from the limited availability of discharge data. It was concluded that the necessary extrapolation of the calibration parameter (bed roughness of main channel) leads to significant uncertainty in simulated design water levels. More intensive measurement of discharges, for which most gated structures are ideal, and a physically more realistic representation of hydraulic structures are self-evident improvements that nevertheless require a cultural shift.

The application of Computational Fluid Dynamics (CFD) in the assessment of flow impact issues that arise long after the start of operation of a structure is rare. Bollaert et al. (2012) employ numerical modelling to assess the influence of gate usage on the formation of plunge pool scour of a hydropower dam. For some issues, like salt water intrusion and sediment transport past a discharge-regulating structure, the solution cannot be found in a modelling tool at one scale. The local flow simulation should in those cases be coupled to a mid- or far-field model that covers a larger area.

A central role nowadays is taken by the multi-disciplinary field of hydroinformatics (Solomatine & Ostfeld 2008, Pengel et al. 2012, Pyayt et al. 2011a, Krzhizhanovskaya et al. 2011, Melnikova 2011), in which different forms of modelling (physics-based and data-driven) are considered and combined with contemporary computational techniques like machine learning (Pyayt et al. 2011b). In the context of the present study, it is noted beforehand that for a complex hydraulic structure, data-driven modelling alone is not an apt option, because a single *Q-H*-relation does not describe all states (Kolkman 1994), or is highly impractical as it would require extensive permanent monitoring.

This study takes the underlying physics as a starting point: elementary flow equations are combined with 2DV time-dependent detail CFD simulations. The method bridges modelling scales with minimum of data coupling and at the same time introduces the use of numerical aids into practical barrier operation for issues that at present are decided upon by expert judgement of the operator.

The remainder of this paper is organized as follows: first, we describe the overall approach, then the method is described in three sections about discharge modelling, free-surface flow simulation and analysis of the modelling results. Next, the results of a series of validation runs for the free-surface model are discussed, followed by results of a test case that gives numerical examples of all modelling steps. We end the paper with recommendations, conclusions and an outlook on future work.

## APPROACH

For obtaining a timely prediction of the flow around gates, we propose a multi-step physics-based modelling strategy which uses data input from a system-scale model. The work-flow of the suggested gate operation system is shown in Figure 1.

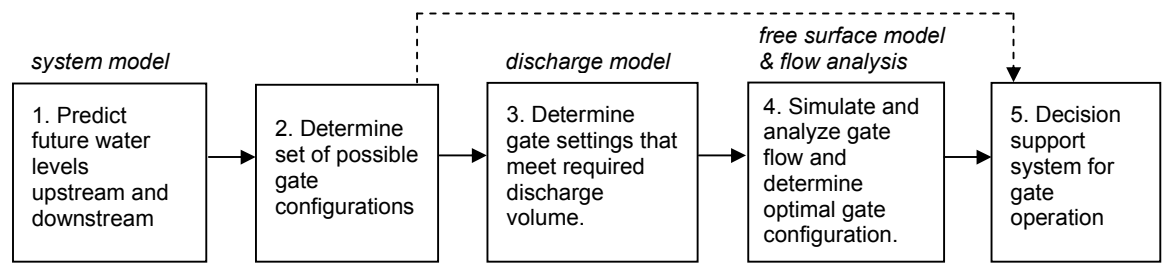

*Figure 1. Scheme of evaluation steps leading to a decision on optimal gate operation. Steps 1-4 are treated in this paper. The dashed line shows the shorter decision sequence taken by barrier systems that do not take into account flow effects.*

The first step consists of the extraction of predicted water levels on both sides of the structure from a far-field (system-scale) model that contains the structure. Different possible gate settings (when to open, how many gates to use) are identified in the second step. All options need to be assessed in terms of discharge capacity; this happens in step 3. In the fourth step of Figure 1, for all gate configurations capable of discharging the required volume, the resulting flow is simulated using CFD. Subsequent analysis of the simulation results determines the impact of the flow for specific issues such as bed stability. The fifth and final step comprises the actual decision of gate operation actions. The conventional sequence of steps taken by most operational systems follows the dashed line in Figure 1, skipping steps 3 and 4. The present study focuses on steps 2-4, which can be seen as an addition to computational decision support systems (steps 1 and 5) by Boukhanovsky & Ivanov (2012) and Ivanov et al.(2012).

A multi-gated discharge sluice with underflow gates will be used to describe the modelling method. After the description of the method, modelling results are presented of validation runs and of an illustrative test case. The central question addressed is how to find the set of gate configurations capable of delivering the required discharge that also meet the relevant constraints on flow properties.

## METHOD

## DISCHARGE COMPUTATIONS

**Configurations of multi-gated structure**

Let us consider the gate configurations of a discharge structure consisting of $n$ similar openings, each accommodating a movable gate. See Figure 2. In its idle state, all $n$ gates close off the openings between the piers and the total discharge is zero. During a discharge event, $m$ gates will be opened partially or completely, allowing a certain discharge through the structure. A 'gate configuration' is defined as the allocation of a number of gates ($m \leq n$) that are opened with a gate opening $a(t)$ while the other gates remain closed. All gates selected for opening will be operated similarly, i.e. with the same $a(t)$.

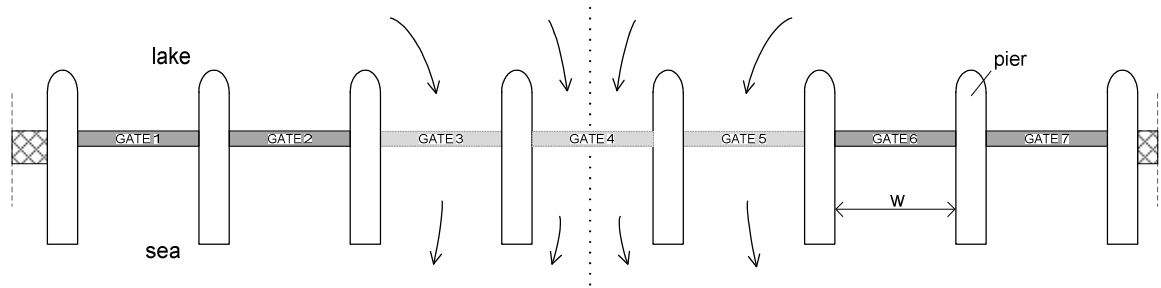

*Figure 2. A multi-gated discharge sluice in plan view. In this example, gates 3, 4, 5 are opened, the others are closed; so n = 7 and m = 3. The dotted line depicts plane of symmetry.*

Before deciding which gates to open, first the possible combinations of opening gates are identified and counted. In general, flow instabilities are not favourable for maintaining an efficient and controllable discharge. As in other parts of physics, symmetry is a global measure for stability of free-surface flows. If asymmetry is allowed, $m$ gates can be chosen freely from the total of $n$ available slots. Then the number of possible combinations is obviously $\binom{n}{m}$, using the common notation for combinatorial choice of $m$ objects out of $n$. For the condition of symmetry to hold, gates may only be opened in such a way that the pattern is symmetric about the vertical plane of symmetry in flow direction (see Figure 2). This implies that the number of options reduces to $\binom{\lfloor n/2 \rfloor}{\lfloor m/2 \rfloor}$ for all $0 \leq m \leq n$, where $m$ cannot be chosen odd if $n$ is even – in which case there are no options at all.

For a structure with seven gates (*n* = 7), for instance, the total number of possible ways to open 1, 2, .., 7 gates is $\sum_{i=0}^{7}\binom{7}{i} = 2^7 = 128$ if asymmetry is allowed and $\sum_{i=0}^{\lfloor 7/2 \rfloor}\binom{\lfloor 7/2 \rfloor}{i} = 16$ if only symmetric configurations are permitted.

This shows that the symmetry constraint greatly reduces the number of ways to open a given number of gates. Furthermore, an even number of gates has roughly half the number of possibilities, because opening any odd number of gates results in asymmetric inflow. This could also hold for an odd-numbered gate structure which misses one (or any odd *m* < *n*) of the gates due to maintenance or operational failure.

**System model and gate control**
The basis is formed by a classic box model, see e.g. Stelling & Booij (1999). The focus is on submerged flow through a multi-gated outlet barrier that blocks seawater from entering the lake at high tide and discharges river water to sea at low tide, see Figure 3. This basic model serves in the present study as a surrogate system-scale model. The water levels it generates will be used as boundary conditions for the near-field modelling.

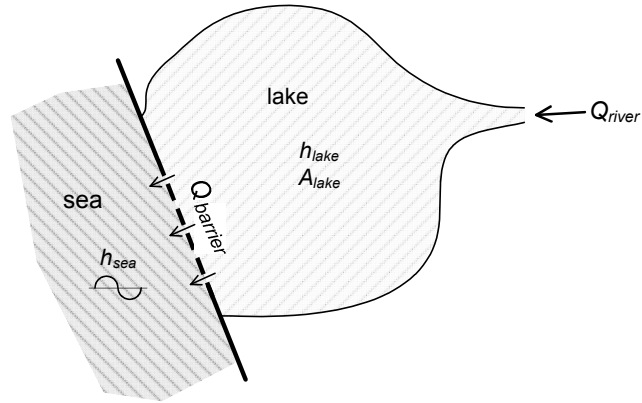

*Figure 3. Classic box model of outflow of a river to sea. An outlet barrier structure regulates the lake level while keeping salt seawater out.*

Assuming barrier gates are closed except when discharging under natural head from lake to sea and assuming zero evaporation, the system is described by:

$$Q_{river} - Q_{barrier} = A_{lake}\frac{dh_{lake}}{dt},$$

where $Q_{river}$ is discharge from a river, $Q_{barrier}$ is the total discharge through gates of barrier, $h_{lake}$ is the water level in the lake, $A_{lake}$ is the area of the lake assumed independent of $h_{lake}$.

Submerged flow past an underflow gate is by definition affected by the downstream water level. The associated discharge depends on both water levels (sea and lake), the gate opening *a* and a discharge coefficient for submerged flow $C_D$. The discharge *Q* through barrier gate *i* is written as

$$Q_i(t) = C_D a_i(t) w \sqrt{2g(h_{lake}(t) - h_{sea}(t))},$$

where *w* is the flow width (see Figure 2) and the subscript "barrier" is dropped from now on. Sea level $h_{sea}$ is approximated by a sine function. The total discharged volume that passes the barrier in the period during which $h_{lake}$ > $h_{sea}$ is found after summing over all *m* gates and integrating with respect to time.

Two gate opening scenarios will be considered. In both scenarios equal gate openings $a(t)$ are applied to all $m$ gates selected for opening. The first scenario uses a constant gate opening $a_{const}$ for the whole discharge period (from $t_{start}$ to $t_{end}$). The opening required to lower the lake level to a desired lake level $h_{target}$ is found by estimating the average required discharge $Q_{tot,req}$ to achieve this and by making estimates of the average discharge coefficient and water levels during the discharge period:

$$a_{const} = \frac{\bar{Q}'_{tot,req}}{m\bar{C}'_D w\sqrt{2g(\bar{h}'_{lake}-\bar{h}'_{sea})}} \text{ with } \bar{Q}'_{tot,req} = \frac{A_{lake}(h_{lake,t(start)}-h_{target})}{t'_{end}-t_{start}},$$

where bars are time-averages and primes indicate predictions of future values. In the second scenario, the discharge is regulated by a Proportional Integral Derivative (PID) controller (Brown 2007). The goal of this scenario is to have a more constant gate discharge by varying the gate openings in time, whilst still achieving the same $h_{target}$ as in the first scenario. The discrete PID formula for discharge at $t_i$ is

$$Q(t_i) = K_P e(t_i) + K_I \sum_{j=1}^{i} e(t_i) + K_D \frac{e(t_i)-e(t_{i-1})}{\Delta t},$$

Where $K_i$ are the gain parameters and the error value is defined as $e(t_i) = Q_{set} - Q(t_{i-1})$. The setpoint $Q_{set}$ is constant and equal to $Q_{tot,req}$, except for linear setpoint ramping applied at the start of discharge to prevent undue fluctuations of gate position. At each time step, the gate opening is derived from this discharge divided by $mC'_D w\sqrt{2g(h_{lake}(t)-h_{sea}(t))}$. Figure 4 shows the flow chart of the system model. It includes computations of the two gate operation scenarios.

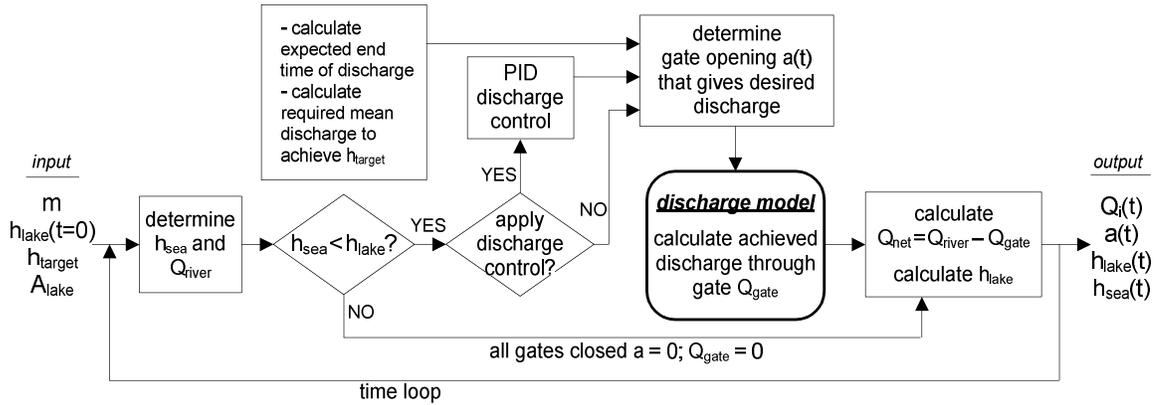

*Figure 4. Flow chart of gate control and water level computations.*

Discharge coefficients actually depend on numerous factors. Also, flows through neighbouring gates influence each other. To distinguish between different gate configurations with the same total flow-through area $m \cdot w \cdot a_{const}$, these two things need to be taken into account. This is done in the discharge model, see the bold block in Figure 4.

**Discharge model**
Vertical lift gates with underflow are raised vertically between piers of a structure. The two main flow types that occur are free flow and submerged flow. When the gates are lifted higher than the water surface, there exists free or submerged Venturi flow (Boiten 1994). All flow types have different discharge characteristics and associated formulae. For estimating the submerged flow discharge, the local water depths are schematized according to Figure 5 (after Kolkman 1994). Conservation of the energy head (Bernoulli equation) is applied in the accelerating parts and the momentum equations in the decelerating parts, yielding a system of four equations (see appendix).

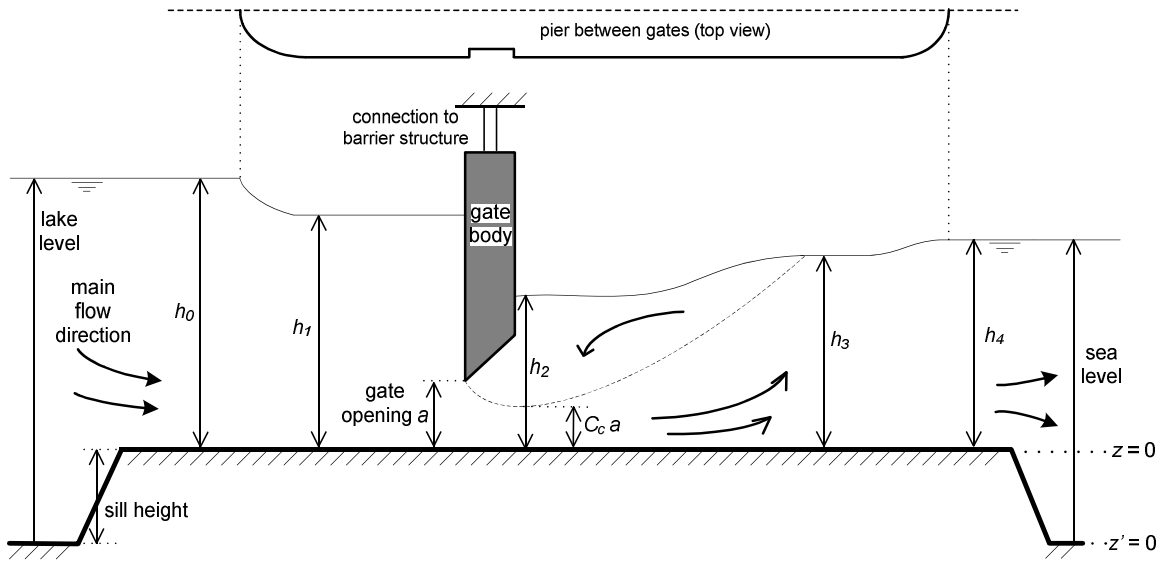

*Figure 5. Definitions of local water depths $h_i$ for underflow gate in hydraulic structure, after Kolkman (1994). Above: top view of pier; below: cross-section free water surface around gate. Sketch not to scale.*

Transitions $h_0$–$h_1$ and $h_3$–$h_4$ with loss coefficients $\xi_{in}$ and $\xi_{out}$ represent the effects of flow entering and leaving the narrow area between two piers. Transitions $h_1$–$h_2$–$h_3$ are the characteristic underflow gate zones, see Battjes (2001) for details. Computations were carried out according to the flow chart shown below with the aim of giving better discharge estimates. The lake and sea levels computed in the system model served as boundary conditions – for variables $h_0$ and $h_4$ of this model, respectively.

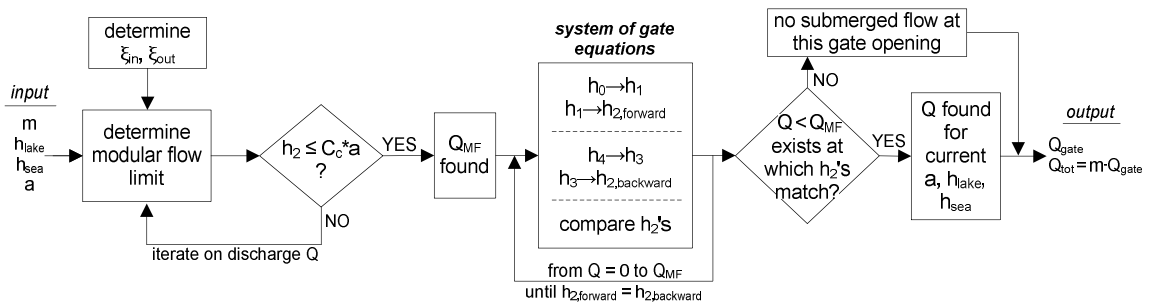

*Figure 6. Flow chart of discharge model. This computation is repeated each time step; it is fully contained in the block named "discharge model" in Figure 4.*

A good geometric design of a discharge-regulator is such that no transition occurs from one flow type to another during regular usage. The model therefore checks if indeed submerged discharge occurs. As criterion for reaching the modular flow discharge $Q_{MF}$, the minimum flow depth in the control section $h_2$ is compared with the flow height in the point of maximum vertical contraction $C_c \cdot a$, the so-called 'vena contracta'. Free and intermediate flow regimes are thus detected, but are not being calculated. Submerged Venturi flow is not considered either, since the idea is to actively *control* the flow.

All four non-linear equations are reshaped into third-order polynomials $f(h_i, h_{i+1}, Q) = 0$. Discharge $Q$ is substituted for the velocity terms and remains as the only unknown in the system of equations. As prescribed for sub-critical flow conditions (Chow 1959), computational direction behind the gate is from downstream to upstream ($h_4$ to $h_2$). On the lake side, computations go in flow direction up to the control sections ($h_0$ to $h_2$). The discharge coefficient $C_D$ is derived from the contraction coefficient $C_c$ for sharp-edged gates, fitted on experimental data cited in Kolkman (1994) so that the full range of gate openings $a/h_1$ is covered, see the appendix for equations.

Iterations on $Q$ ultimately yield a value at which $h_{2,forward}$, computed from upstream, is equal to $h_{2,backward}$ computed from downstream. This is the achieved value of $Q$ for the given gate opening $a$. The entrance and exit losses are assumed to depend on the number of gates in use ($m$). The method does not distinguish between different gate configurations with equal $m$, however. Numerical results are shown in the results section.

## CFD SIMULATIONS

Step 4 in Figure 1 consists of two parts: free-surface CFD simulations (discussed in this chapter) and flow analysis (discussed in the next chapter).

### Model set-up

A non-hydrostatic flow model is applied to find out which of the selected gate settings is most favourable in terms of flow properties. The two-dimensional domain (2DV) is defined by a vertical cross-section through the gate section from one water body (lake) to the other (sea), see Figure 7. A rigid rectangular gate with a sharp-edged bottom is modelled implicitly by cutting its shape out of the flow domain. The Reynolds-Averaged Navier-Stokes (RANS) equations for incompressible flow, included in the appendix, are the basis for the simulations. Figure 7 gives the flow chart of the CFD simulations. The model domain covers the flow from $h_1$ to $h_3$. These input values are taken from the discharge model.

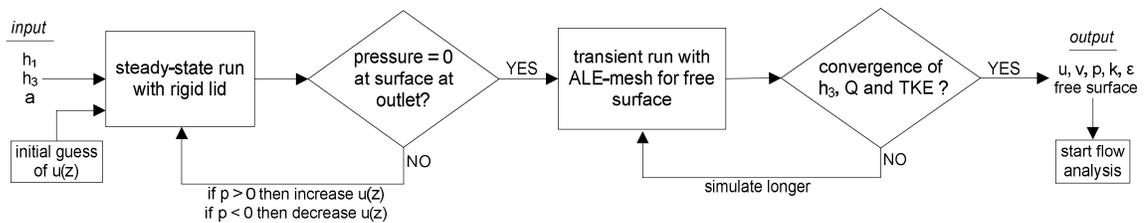

Figure 7. Flow chart of FEM free-surface flow simulations.

For each simulated flow situation two consecutive runs are made: a steady-state run and a time-dependent transient run. In the former run, iterations on the outflow velocity profile are done until pressure at the surface becomes zero. The results of this pre-run are then implemented as initial conditions for the transient run, which uses a moving mesh to simulate the free surface. Boundary conditions are similar for both runs except for the surface downstream of the gate, see Figure 8.

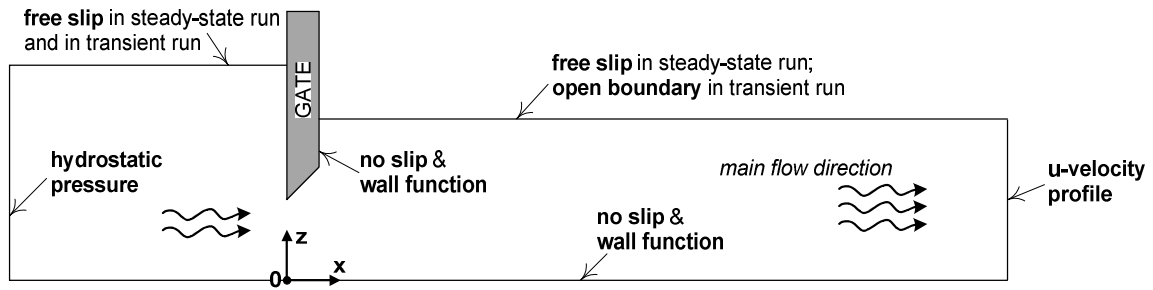

Figure 8. Boundary conditions of CFD model. The main flow direction is from left to right. Sketch not to scale.

The upstream flow boundary consists of a hydrostatic pressure profile $p(z) = \rho g(h_1 - z)$. The downstream boundary is a block profile $u$-velocity. No slip is applied at the walls ($\vec{u} = 0$) along with a wall function. The steady pre-run uses a 'rigid lid' (free slip boundary, $\vec{u} \cdot \vec{n} = 0$) for the downstream water surface. The upstream free surface is modelled as a rigid lid in both runs.

An unstructured computational mesh is used with refinements near the bottom wall and gate boundaries, made up of around 35,000 triangular elements and yielding about 230,000 degrees of freedom for a transient run. Figure 9 shows part of the mesh. The Arbitrary Langrangian-Eulerian (ALE) method with Winslow smoothing (Donea et al. 2004) is applied to compute the deformation of the computational mesh downstream of the gate. At the top boundary in the transient run, the velocity condition is an open boundary with zero stress in normal direction. At the same boundary the mesh velocity in normal direction is prescribed as $u_{mesh,n} = u \cdot n_x + w \cdot n_z$ (Ferziger & Perić 2002). Mesh convergence tests showed that the applied mesh is sufficiently dense so that results do not improve on further mesh refinement.

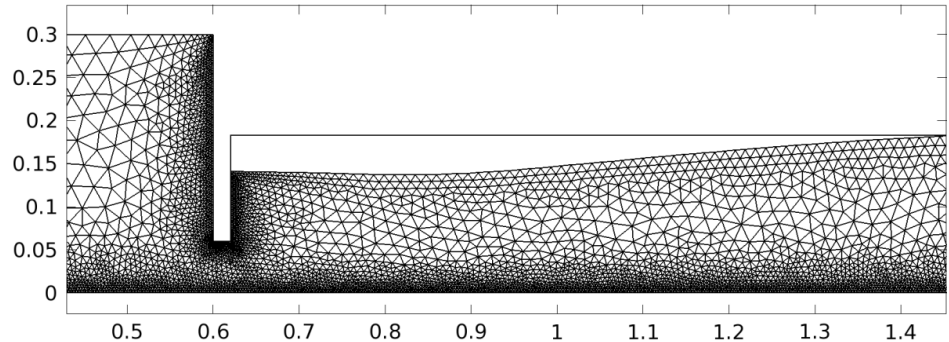

*Figure 9. Example snapshot of detail of computational mesh. Deformed surface downstream of gate is visible. Flow is from left to right.*

The more common choice of applying a velocity condition upstream and a pressure boundary downstream conflicts with the required ALE moving mesh condition at the outlet boundary. Vertical mesh freedom is necessary for the surface movement. A hydrostatic pressure profile cannot be prescribed at the outlet, since any change in water depth at this boundary would imply a change of local pressure, which contradicts the applied pressure profile.

In the course of the transient run, the freesurface adapts to the pressure field and vice versa. Because the physical flow situation is quasi-steady, with fluctuations depending on degree of submergence and gate opening, the surface may show oscillations in time in its equilibrium state. As a consequence, the flow discharge is also not strictly constant in the equilibrium state.

The package Comsol Multiphysics is used to simulate the gate flow. This Finite Element Method (FEM) solver is applied to solve the discretised RANS equations. The generalized alpha time-implicit stepping method is applied to ensure Courant stability, with a strict maximum time step of $\Delta t = 0.02$s. The variables are solved in two segregated groups using a combination of the PARDISO solver and the iterative BiCGStab solver in combination with a VANKA preconditioner. The standard k-epsilon model is used for turbulence closure. Simulation of 24 seconds of physical time took around six hours of wall-clock time on Intel 8-core i7 processor, 2.93 GHz, 8 Gb RAM, occupying on average 1 Gb RAM and 50% of total CPU power.

## ANALYSIS

The second part of step 4 in Figure 1 is the analysis of the modelling results obtained in previous steps. In this chapter, three aspects of analysis are discussed: flow parameters, vibrations and bed stability.

**Flow parameters**

Three parameters that are required for assessing various types of flow impact are extracted from the CFD model: the contraction coefficient $C_c$, the velocity in the vena contracta $U_{vc}$ and the Froude number *Fr*. The flow field is interpolated to a regular grid, so that the edge of the separated layer is found, see Figure 13. The contraction coefficient is thus found directly.

The cross-sectional averaged velocity in the vena contracta is defined by a spatial average in the separated shear zone:

$$U_{vc}(t) = \frac{1}{C_c(t)a} \int_{z=0}^{C_c(t) \cdot a} U(z,t) \, dz$$

where *U* is the velocity magnitude scalar at the point of maximum flow contraction. For quasi-steady gate flow with significant fluctuations, the temporal mean of this quantity, $\bar{U}_{vc}$, may be used.

The Froude number is a widely used dimensionless measure for surface disruption. It is of use for finding the transitions to intermediate and free flow regimes and predicting modular flow discharge and associated gate opening. Here it is defined as

$$Fr(t) = \frac{U_{vc}(t)}{\sqrt{gh_2(t)}}$$

in which obviously $h_2 = C_c \cdot a$ in fully free flow. An overview of critical flow theory from a historical perspective is given by Castro-Orgaz & Hager (2010) and from a more practical viewpoint by Boiten (1994). In a more complete flow assessment, not only the contraction caused by the vertical gate is used as a criterion for modular flow, as is done here, but also contraction caused by horizontal and possibly vertical flow domain transitions at the inlet of the structure should be included.

**Vibrations**

The interaction of current with the movable hydraulic gate is capable of causing significant flow-induced vibrations (FIV). Although dedicated design tests greatly reduce susceptibility for dangerous dynamic forces, active prediction and control will broaden the windows of structure operation.

Literature on dynamic gate forces caused by this phenomenon uses a dimensionless parameter of reduced velocity to signify occurring gate vibrations (Hardwick 1974, Billeter & Staubli 2000, Erdbrink 2012). In time-dependent form it is written as

$$Vr(t) = \frac{U_{vc}(t)}{f_{gate}(t) \cdot L}$$

Where $f_{gate}$ is the response frequency of the structure in Hz; *L* is a characteristic length scale of the gate, usually the thickness of the gate bottom, and $U_{vc}$ as defined in the previous section. The response frequency is not easily determined analytically (see general formula in appendix); among other reasons because the 'added' water mass $m_w$ that is caused by the inertia of water being pushed away by the gate deviates from analytical values at non-zero gate flow (Blevins 1990). The gate frequency may best be monitored in situ by installing sensors – which ought to be sensitive to small amplitudes in order to have predictive value. Erdbrink et al. (2012) provide a recipe for a data-driven gate control system for gate vibrations. It is therein proposed to combine physics-based modelling and sensor data with machine learning computations to steer the gates clear of riskful situations.

From numerous experimental studies (Naudascher & Rockwell 1994) it is concluded that for one specific gate, the amplitude *A* due to FIV, in cross-flow or in-flow direction or both, is a function of *Vr*, *a* and submergence:

$$A = f(Vr, a, h_3)$$

Details of the gate geometry are decisive for occurrence or absence of vibrations. A database with response data from past laboratory studies could be used to predict amplitudes of future flow situations in an operational system.

**Scour and bed protection**

The prediction of local scour downstream of weirs and sluice structures caused by outlet currents is described by Breusers (1966) and Hoffmans & Pilarczyk (1995). In their design formulae they use turbulence parameters to predict the depth of the scour hole and in unprotected beds. For beds protected with granular material (loose rocks), the Shields parameter is a classic non-dimensional measure applied as a first indicator for instability (Shields, 1936). An adapted version of this parameter used by Jongeling et al. (2003) and elaborated upon by Hofland (2005) and Hoan et al. (2011) is defined as

$$\Psi(x) = \frac{\langle (\bar{U}(x) + \alpha\sqrt{k(x)})^2 \rangle}{\Delta g d(x)} \quad \text{with} \quad \Delta = \frac{\rho_s - \rho_w}{\rho_w},$$

where $\langle .. \rangle$ denotes spatial averaging over the whole water depth, $k$ is the turbulent kinetic energy (TKE), $d$ is the local water depth, $\bar{U}$ is the mean flow velocity magnitude and $\alpha$ is an empirical parameter for bringing into account the turbulence that depends on flow type and local geometry (e.g. slopes in bottom profile).

## MODEL VALIDATION RESULTS

A series of validation runs was done for the free-surface model. 'Validation run' is used here in the meaning discussed by Stelling & Booij (1999): the uncalibrated model is run without any tweaking of parameters to see if it can reproduce the most important physical features. Experimental laboratory data by Nago(1978,1983) for a vertical sharp-edged gate under submerged efflux serve as comparison. Nago's(1978,1983) dimensions were used without any scaling. His discharge formula doesn't contain the downstream level $h_3$ explicitly. Its influence is instead found in the discharge coefficient $C_E = Q/(aw\sqrt{2gh_1})$. The simulated discharge is computed by spatial integration of horizontal velocity at the outflow boundary. In Figure 10 coefficient $C_E$ is plotted for different series of dimensionless gate openings and for a range of dimensionless downstream levels.

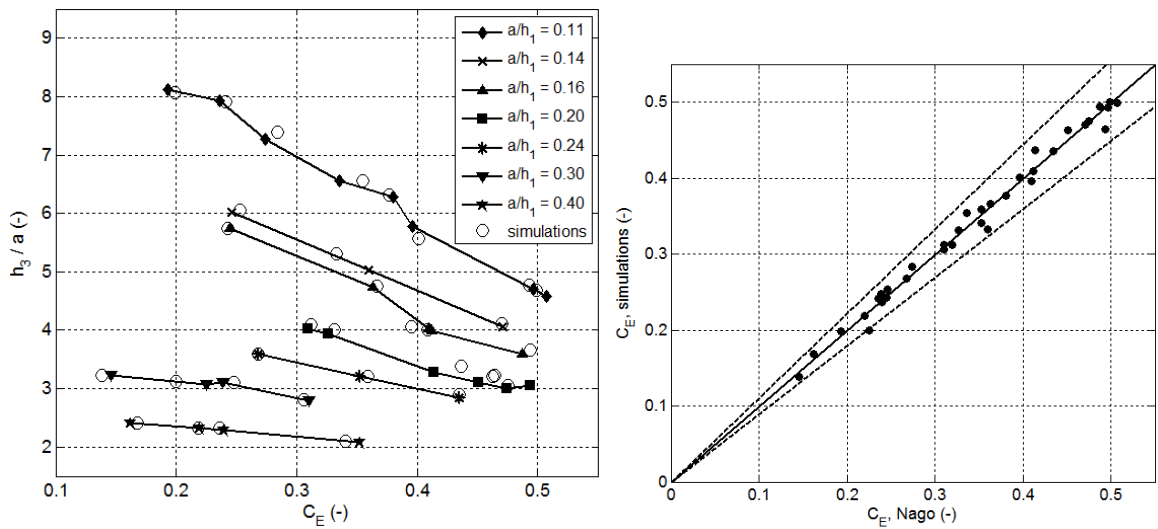

*Figure 10. Results of validation runs showing discharge coefficient $C_E$ simulated by the free-surface CFD model versus experimental data of submerged flow of a sharp-edged underflow gate by Nago (1978, 1983). Left: Sorted by gate opening ($a/h_1$) and downstream level ($h_3/a$). Right: direct comparison of the same data. Dashed lines mark 10% deviation.*

The results of the validation runs make clear that the simulations capture the discharges of the experimental data quite accurately: the correlation coefficient is 0.994 and the root mean square error is 1.14%. The fact that the uncalibrated model shows good discharge estimates gives confidence in the predictive power of this modelling approach. Physical output not validated here (such as TKE) may be calibrated in future studies by adjusting suitable model parameters. Convergence of various flow variables occurs at different rates. First, the mean velocities stabilize, and then the forces on the gate converge, then the discharge, and lastly the turbulent energy.

The chosen boundary conditions proved to lead to stable results for all submergence ratios of Nago's (1978, 1983) data. It was found that the moving mesh is the critical factor for numerical stability. ALE is a suitable method for computing the free surface for quasi-steady gate flow as long as the flow remains submerged. Steep surface gradients associated with lowering $h_3$ cause inverted mesh elements and hence numerical instabilities.

## TEST CASE RESULTS

The described methods are illustrated by a test case example. The results of three modelling steps are discussed: the sluice model containing the system model (for water levels) plus the discharge model (Figures 4 and 6), the free-surface model (Figure 7) and analysis of vibrations and bed stability. Four tidal cycles and four discharge events were modelled for a discharge sluice with seven gates regulating a lake with constant river inflow. The goal of the computations is to determine the optimal number of gates to open and the best gate operation scenario.

### Results of system and discharge model
*Model parameters*
- $n = 7$, $m = 1, .., 7$
- $A_{lake} = 1.9 \cdot 10^7$ m$^2$
- $Q_{river} = 100$ m$^3$/s
- $h_{lake}(t=0) = 6.1$ m
- $h_{target} = 6.0$ m
- $w = 22.5$ m
- sill height: 3 m
- mean sea level = $z'$ + 6.1 m
- tidal amplitude = 0.60 m
- tidal period = 12.5 hours

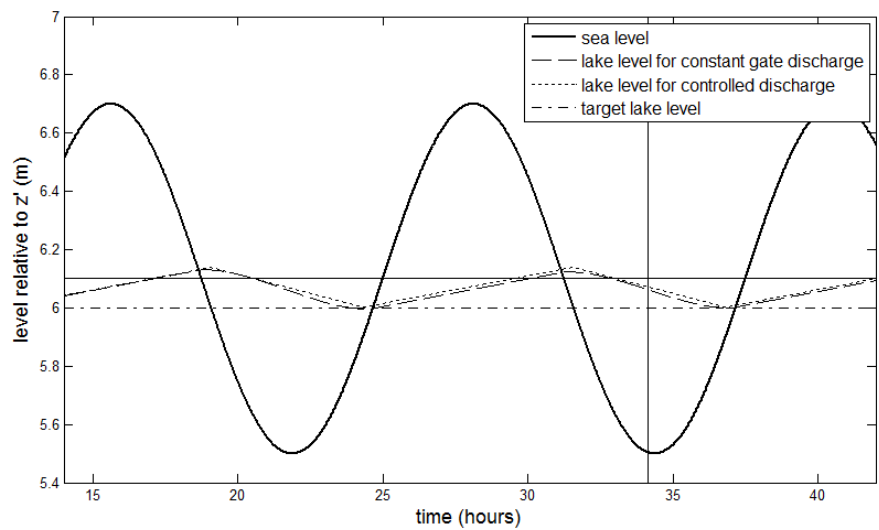

*Figure 11. Results of sluice model for $3 \leq m \leq 7$: sea and lake level for gate operation scenario with and without PID-controlled discharge.*

The sluice model was run for $1 \leq m \leq 7$. When opening only one gate, the target lake level could not be reached even when lifting the gate completely. When using two gates, the target level is reached, but the modular flow limit is exceeded for the greatest part of the discharge period. This results in unwanted transitions to intermediate and free flow with fluctuating discharges that are hard to control. For $3 \leq m \leq 7$ strictly submerged flow exists and the target is met. Therefore, only these configurations are modelled further. The plotted water levels (Figure 11) show that the lake level fluctuates in a controlled way and nearly identically for the scenarios with and without discharge control.

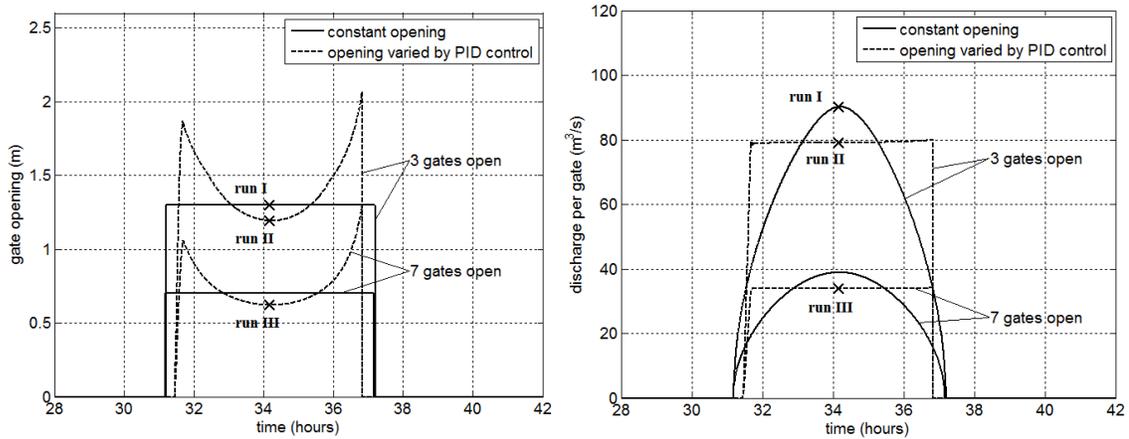

*Figure 12. Results of sluice model: gate openings (left) and achieved discharges per gate (right).*

In Figure 12, the gate openings and achieved gate discharges in time are plotted for one tidal period for the situations with three or seven gates opened during the discharge event. Intermediate numbers of operated gates ($4 \leq m \leq 6$) lie between the shown curves for $m = 3$ and $m = 7$, but are not plotted for clarity. It can be seen that constant gate openings give discharges that vary in time following the time-dependent hydraulic head difference. In the PID-controlled scenario, the gate opening is automatically operated in such a way that the discharge stabilizes quickly after the start. Naturally, the areas under the Q-graphs, equal to the total discharged volume $V_{tot}$, are roughly the same regardless of gate opening scenario.

The globally achieved mean discharge coefficient $\bar{C}_D$ of the discharge event of the previous tidal period is computed by the model and used to improve the prediction of required gate opening for the next discharge event.

From these results, three configurations are selected for evaluation by free-surface simulations. These cases are marked in Figure 12 as runs I, II and III. Runs I and III represent extremes: a constant gate opening with only three gates in use (high Q) and a controlled opening with all seven gates in use (low Q). All three runs are at the time of maximum head difference. In real-life practice, more cases could be selected for simulation depending on specific interests and available computing power.

**Results of CFD simulations**

To simulate the two selected runs I and II within the validated range, the levels and opening are scaled down with length scale 1:10, see Table 1.

*Table 1. Values of selected CFD runs.*

| run | gate configuration | length scale | $h_0$ (m) | $h_1$ (m) | $h_3$ (m) | $h_4$ (m) | gate opening a (m) | total discharge $Q_{tot}$ (m³/s) | discharge per gate $Q_i$ (m³/s) | discharge per gate per unit width $q_i$ (m²/s) |
|---|---|---|---|---|---|---|---|---|---|---|

| | | | | | | | | | |
|---|---|---|---|---|---|---|---|---|---|
| I | m = 3, constant opening | 1:1 | 3.06 | 2.95 | 2.41 | 2.50 | 1.30 | 270 | 90.2 | 4.01 |
| | | 1:10 | 0.306 | 0.295 | 0.241 | 0.250 | 0.130 | 0.855 | 0.285 | 0.127 |
| II | m=3, PID control | 1:1 | 3.07 | 2.99 | 2.43 | 2.50 | 1.19 | 237.61 | 79.20 | 3.52 |
| | | 1:10 | 0.307 | 0.299 | 0.243 | 0.250 | 0.119 | 0.751 | 0.250 | 0.111 |
| III | m = 7, PID control | 1:1 | 3.07 | 3.06 | 2.49 | 2.50 | 0.622 | 238 | 33.96 | 1.51 |
| | | 1:10 | 0.307 | 0.306 | 0.249 | 0.250 | 0.0622 | 0.752 | 0.107 | 0.0477 |

■ input values for CFD runs.
All water levels $h_i$ are relative to $z = 0$.

The near-gate flow velocities, pressures, TKE and dissipation are simulated. Figure 13 shows a plot of the simulated flow field (at length scale 1:10) of run II by indicating ($u,w$)-vectors.

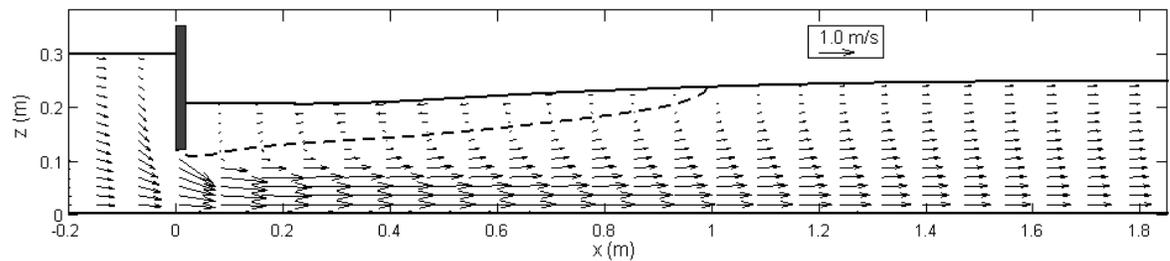

*Figure 13. Vector flow field of run II. Flow is from left to right. The computed free surface behind the gate shows local lowering. Dashed line indicates separation between positive and negative u-velocities. The figure shows only part of the actual computational domain. Total domain length is 3.6 m.*

The simulated free surface as expected sinks in the region directly downstream of the gate (solid line in Figure 13). In this case, the vena contracta is located at short distance downstream of the flow separation point. The separation between positive and negative horizontal velocities in the recirculation area is derived (dashed line in Figure 13). At a distance of around five times the downstream water level past the gate, the flow reattaches at the surface and the velocity starts to return to a more uniform profile.

Figure 14 shows plots of the pressure and turbulent kinetic energy of run II.

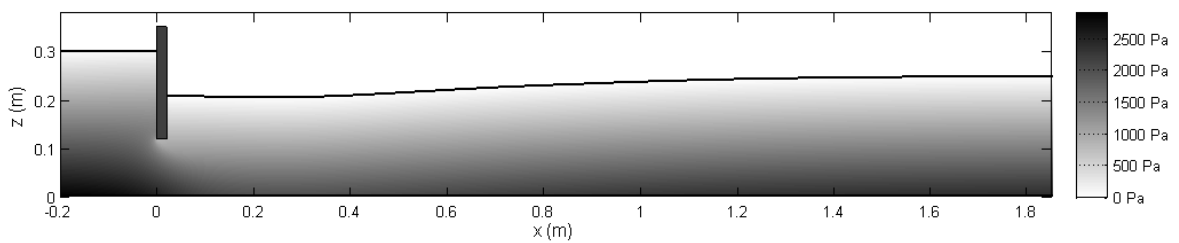

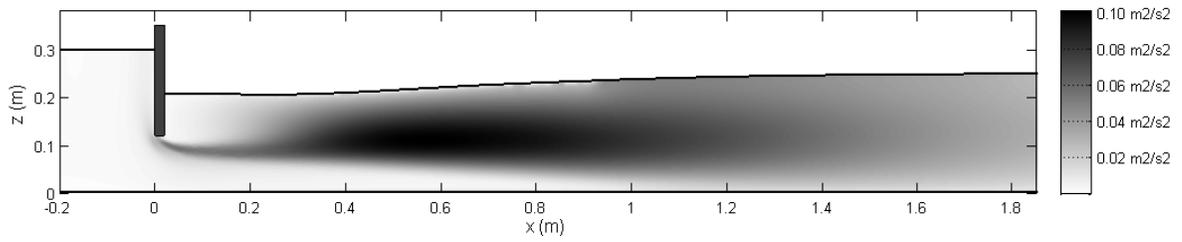

*Figure 14. Pressure p in Pa (above) and turbulent kinetic energy k (TKE) in m²/s² (below) of run II.*

In the case shown in the plots, the equilibrium state reached in the simulations is fully steady. Pressure gradients are mild; the pressure returns smoothly to a hydrostatic shape as the streamlines become parallel downstream. The TKE reaches a maximum in the middle of the water column at about two times the downstream water depth past the gate. Run I has a steeper surface behind the gate than run II (shown in Figure 13 and 14) and higher TKE levels, while run III has the lowest TKE levels and the most level surface downstream of the gate.

**Results of flow analysis**

The output of the CFD free-surface model is used for computing the values of the three flow parameters that were discussed in an earlier section, see Table 2.

*Table 2. Computed flow parameters derived from CFD model results*

| run | $C_c$ (-) | $U_{vc}$ (m/s) | Fr (-) |
|---|---|---|---|
| I | 0.86 | 3.37 | 0.76 |
| II | 0.90 | 3.35 | 0.74 |
| III | 0.86 | 2.54 | 0.52 |

Table 2 shows that the contraction coefficients do not differ much, which is expected for similar gate types. The velocity in the control section $U_{vc}$ is highest for the situation with highest discharge per gate (run I) and lowest for the situation with smallest discharge per gate (run III). The same holds for the Froude number. This matches observations from the free surface curvatures of the final solution of the transient simulations.

The flow impact on the bed protection material is estimated by computing $\Psi$ for two different α for the selected runs. The whole water depth $d$ is used for averaging the square of the maximum local velocity term $(\bar{U} + \alpha\sqrt{k})^2$. The results are plotted in Figure 15.

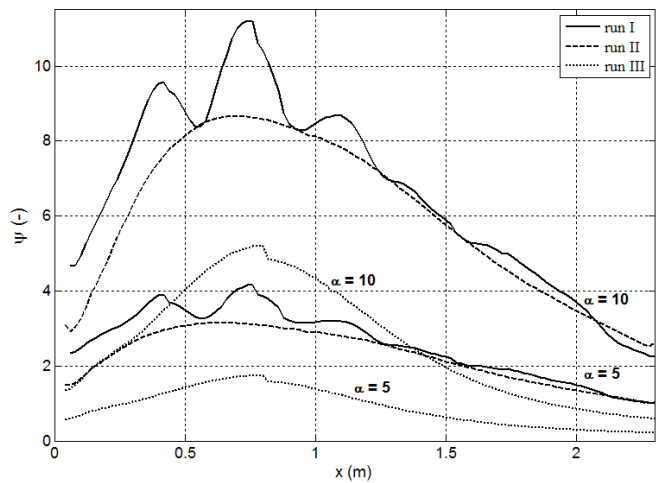

*Figure 15. Computed values of bed stability parameter Ψ downstream of the gate for two different values of turbulence impact parameter α. Runs I, II and III are shown.*

The plot shows that run I (three gates with constant opening) has the strongest flow impact on the bed material of the three runs irrespective of the choice for α. The pronounced surface curvature closely behind the gate results in a number of peaks in the Ψ curve of run I. The Ψ–values of run II show that controlling the discharge without opening more gates already has a lower, smoother and hence more predictable flow impact on the bed. Run III (seven gates with controlled discharge) has the lowest flow impact. All runs reach their maximum flow impact on the bed around the same (limited) distance downstream of the gate. For all runs the general shape of the curves is quite similar for both values of α, indicating that turbulence is dominant over mean velocity for the flow impact.

Overall the values of the bed stability parameter are somewhat low compared to previous numerical investigations by Erdbrink & Jongeling (2008) and Erdbrink (2009), which could be attributed to the use of the standard k-epsilon model in this study instead of the RNG k-epsilon turbulence model used in the two mentioned studies. Choosing higher α values could compensate the lower TKE. For practical application one should fix α after calibration in experimental investigations and one should define a threshold value for Ψ not to be exceeded during operation to be used for measuring the fitness of different flow scenarios.

Turning to the assessment of gate vibrations, it is calculated that for an assumed range of structural response frequencies of 2–5 Hz (typical values for large hydraulic gates), the reduced velocity number $V_r$ lies in the range 3.5–8.5 for runs I and II and in the range 2.5-6 for run III. For illustration purposes, a response curve is devised, see Figure 16, since a full evaluation is rather laborious (e.g. Billeter & Staubli 2000). Projection of the $V_r$-values onto the response curve give resulting vibration amplitudes.

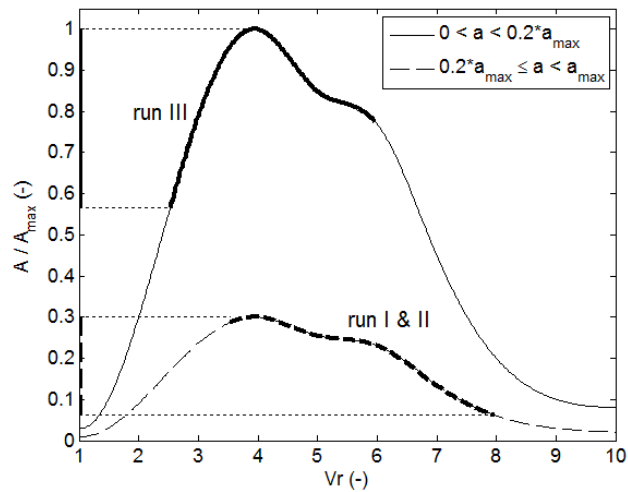

*Figure 16. Gate vibration response for runs I-III giving relative amplitude $A / A_{max}$ as a function of reduced velocity $V_r$. Fictitious response curves are used to illustrate the method. Two regions of gate openings* a *are distinguished.*

Two different response curves are used in Figure 16. The most significant excitation of cross-flow vibrations of a flat-bottom gate occurs at small gate openings, therefore higher amplitudes are expected for run III. In this fictitious case, the computed $V_r$-ranges indeed give higher relative amplitudes for run II than for the other two runs. As with the bed protection assessment, the definition of a literature-based threshold level would be a logical addition for real applications.

Based on the discussed modelling results and flow analysis, it may be decided to implement the discharge scenario of run II, because it leads to acceptable vibration levels and gives a lower impact on the bed material than run I – while still ensuring sufficient discharge volume to reach the target lake level.

## RECOMMENDATIONS

As a main recommendation, we propose to apply this modelling process in a case study of existing barrier structures, such as Haringvliet, Oosterschelde, Maeslantkering in The Netherlands and the Saint Petersburg barrier in Russia. This research should find a natural place within on-going work on system-scale modelling for water level prediction used in decision support for hydraulic structures (Boukhanovsky & Ivanov 2012).

Specific gate uses are to be simulated and evaluated. For the last two barriers just mentioned, the operational modelling system will be mostly aimed at widening the window of operation. The introduced methods can also be adapted for weirs in rivers. Coupling the presented models with a mid-field or far-field model of a regional model would enable an operational impact assessment for water management issues such as salt water intrusion.

The inclusion of measurement data (from field sensors or laboratory tests) is necessary for the calibration of empirical parameters (such as entrance and exit losses), for the process of model validation and for providing actual model input (water levels). Experiences from the field of hydroinformatics should be added to the present research to make the extension towards data-driven modelling components. The link with data assimilation that is to be accommodated by the higher-level models is obvious.

A longstanding issue in the engineering practice of detailed hydrodynamics is turbulence modelling. The right balance between accuracy and computational costs needs to be found for specific applications. Again, smart use of measurement data for numerical validation and calibration could be the key. It is furthermore expected that intermediate and free flow conditions where hydraulic jumps occur away from the gate can be modelled more universally using other numerical methods such as Phase Field or Volume Of Fluid. If needed the model can thus be extended to account for dynamic effects directly related to opening and closing actions of the gates. Active setpoint ramping of the PID-control using feed-forward model predictions is another recommendation related to this.

## CONCLUSIONS AND FUTURE WORK

The purpose of the current study was to set up physics-based modelling methods for a flow-centred operation of gates of hydraulic structures. The described case of a multi-gated outlet barrier sluice has shown how discharge estimates and free-surface simulations can aid in deciding on optimal gate configuration and opening scenarios.

The application of a PID-controller to achieve a more constant discharge during changing head differences emerged as a feasible addition to traditional structure operation. Prediction of gate discharge coefficients is a central issue in determining appropriate gate openings in the control process. A combination of elementary equations and empirical relations was used for this. Increase in computational power over the years now enables solving these flow equations in quick assessment procedures during operation.

Free-surface Computational Fluid Dynamics simulations of the turbulent flow past an underflow gate revealed the effects of local lowering of the surface on flow velocities and turbulent kinetic energy levels. Time-dependent FEM simulations with a moving mesh technique were found to give stable solutions of the free-surface under submerged conditions. From a series of validation runs it is concluded that the free-surface model yields discharge values for a range of gate openings and submergence levels within an acceptable accuracy of experimental values.

Among the flow analysis possibilities based on output from the free-surface model is computation of the Froude number, the reduced velocity parameter for estimating gate vibrations and a stability parameter for granular bed protection. The numerical example of the discharge sluice has proved the feasibility of combining discharge estimates with free-surface simulations for deriving operational decisions. For the particular case treated in this paper, it was found that lower turbulent kinetic energy levels of the PID-controlled discharge scenarios contribute significantly to reducing the flow attack on the bed protection. Additionally, the model showed the influence of the number of opened gates on the flow properties.

The practical benefits of including near-field flow modelling in gate control systems seem clear. It will enable more sophisticated water reservoir management in everyday operation with respect to issues such as salt water intrusion, fish migration and possibly saving energy. In extraordinary situations, model results can help maintain safe gate usage and prevent gate vibrations, washing away of bed protection and the development of scour holes around the structure.

Limitations of the followed modelling approach need to be addressed in follow-up studies. Additional calibrations are necessary: PID-control optimization to obtain the desired discharge more precisely, discharge and loss coefficients in the flow equations and turbulence model parameters. Next to this, improvements to the free-surface model should broaden the range of applicability so that steeper surface disruptions and hydraulic jumps as found in free flows can be captured as well.

The physics-based model of this study is logically complemented by data-driven techniques in future studies. It is believed that hydroinformatics provides the required tools for this. Use of sensor data from real-life structures and coupling to system-scale water level prediction models are seen as next steps. Moreover, it should be investigated how operational decisions should be derived when taking into account the various criteria and flow constraints.

**Acknowledgements**
This work is supported by the EU FP7 project *UrbanFlood*, grant N 248767; by the *Leading Scientist Program* of the Russian Federation, contract 11.G34.31.0019; and by the *BiG Grid* project BG-020-10, #2010/01550/NCF with financial support from the Netherlands Organisation for Scientific Research NWO. It is carried out in collaboration with Deltares.

## APPENDIX: FORMULAS

(1). Combining the empirical graphical formulations of Henry (1950) and Cozzo (1978), the contraction coefficient for sharp-edged gates is assumed equal to:

For $\frac{a}{h_1} > 0.5$ (Henry regime), $C_c = \frac{0.782}{1.782 - a/h_1}$ ;

For $\frac{a}{h_1} \leq 0.5$ (Cozzo regime), $C_c = -0.004 \log(a/h_1) + 0.6074$

(2). Relation discharge coefficient and contraction coefficient

$$C_D = \frac{C_C}{\sqrt{1 + C_C a/h_1}}$$

(3). Underflow gate equations
The equations, denoted [$h_i$, $h_{i+1}$], describe the transition from water level $h_i$ to $h_{i+1}$ ($h_{i+1}$ being located downstream of $h_i$), see Figure 5:

[$h_0$, $h_1$]
$$h_0 + \left(\frac{Q}{w_0 h_0}\right)^2 / 2g = h_1 + (1 + \xi_{in}) \left(\frac{Q}{w h_1}\right)^2 / 2g$$

[$h_1$, $h_2$]
$$h_1 + \left(\frac{Q}{w h_1}\right)^2 / 2g = h_2 + \left(\frac{Q}{w C_c a}\right)^2 / 2g$$

[$h_2$, $h_3$]
$$\frac{1}{2} \rho g w h_2^2 + \frac{\rho Q^2}{w C_c a} = \frac{1}{2} \rho g w h_3^2 + \frac{\rho Q^2}{w h_3}$$

[$h_3$, $h_4$]
$$h_3 - h_4 = \frac{U_4^2}{2g} + (\xi_{out} - 1) \left(\frac{Q}{w h_3}\right)^2 / 2g$$

$$\xi_{in} = \left(\frac{1}{C_{c,in}} - 1\right)^2$$

(4). RANS equations for incompressible flow under gravity (transient form)

$$\rho \frac{\partial \vec{U}}{\partial t} + \rho \vec{U} \cdot \nabla \vec{U} + \nabla \cdot \overline{(\rho u' \otimes u')} = -\nabla P + \nabla \cdot \mu \left(\nabla \vec{U} + (\nabla \vec{U})^T\right) + \rho g$$
$$\rho \nabla \cdot \vec{U} = 0$$

where Reynolds decomposition is defined by $\vec{u} = \vec{U} + u'$, $\otimes$ is the outer product, $\mu$ is the dynamic viscosity in Pa·s. The steady-state form follows from $\partial \vec{U}/\partial t = 0$.

(5). Discharge coefficient used by Nago (1983)

$$C_E = \frac{Q_{gate}}{aw\sqrt{2gh_1}} = C_D \frac{\sqrt{h_1 - h_3}}{\sqrt{h_1}}$$

(6). Vibration frequency of partly submerged gate

$$f = \frac{1}{2\pi} \sqrt{\frac{k + k_w}{m + m_w}}$$

With $k_w$ the added rigidity of the system due to Archimedean force on the gate body and $m_w$ the added water mass. Damping is neglected in this formula.